\documentclass[11pt]{cernrep}
\usepackage{graphicx}
\usepackage{cite,./mcite}
\usepackage[latin1]{inputenc}
\usepackage[dvips]{epsfig,color}
\usepackage{wrapfig,rotating}
\usepackage{amssymb,amsmath,array}

\newcommand{\spom} {\mbox{$\scriptstyle \mathrm{I}\! \mathrm{P}$}}

\newcommand {\lapprox}
   {\, \raisebox{-0.7ex}{$\stackrel {\textstyle<}{\sim}$} \,}

\begin{document}
\title{Towards a Combined HERA Diffractive Deep Inelastic Scattering 
Measurement\footnote{ \ \ Contributed to the Proceedings of the HERA-LHC
Workshop \cite{heralhc}.}}
\author{Paul Newman$\,{}^a$, Marta Ruspa$\,{}^b$}
\institute{
${}^a$ School of Physics \& Astronomy, University of Birmingham, B15 2TT, UK. \\
${}^b$  Universit\`a del Piemonte Orientale, 28100 Novara, Italy. \\
}
\maketitle
\begin{abstract}
The diffractive 
dissociation of virtual photons, $\gamma^{\star}p \to Xp$, has been studied 
with the H1 and ZEUS detectors at HERA 
using various complementary techniques. Events have been selected
by direct 
tagging of the outgoing proton or by requiring a large 
rapidity gap between the proton and the system $X$.
The diffractive contribution has also been unfolded by decomposition of the
inclusive hadronic final state invariant mass distribution. Here,
detailed comparisons are made
between diffractive cross section measurements obtained 
from the different methods 
and the two experiments, showing them to be consistent
within the large uncertainties associated with the treatment of
proton dissociation processes. First steps 
are taken towards the combination of the H1 and ZEUS results.  
\end{abstract}

\section{Introduction}

\begin{wrapfigure}{r}{0.25\columnwidth}
\vspace{-0.3cm}
\centerline{\includegraphics[width=0.25\columnwidth]{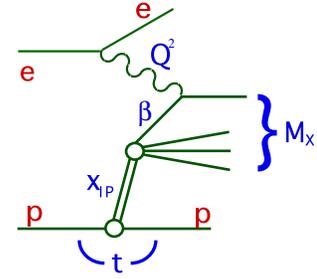}}
\caption{Illustration of the kinematic variables describing the
virtual photon dissociation process, $\gamma^{\star}p \to Xp$, 
in $ep$ collisions.}
\label{fig:diagram}
\end{wrapfigure}
In the single diffractive dissociation 
process in proton-proton scattering, $pp \rightarrow Xp$, 
at least one of the beam hadrons emerges intact from 
the collision, having lost only a small fraction of its energy 
and gained only a small transverse momentum. 
In the analogous process involving 
virtual photons, $\gamma^{\star}p \to Xp$ 
(figure~\ref{fig:diagram}) \cite{zeus:obs,h1:obs}, 
an exchanged photon of virtuality $Q^2$ 
dissociates through its interaction with the proton at a 
squared four momentum transfer $t$ 
to produce a hadronic system $X$ with mass $M_X$.
The fractional longitudinal
momentum loss of the proton during the interaction is
denoted $x_{\spom}$, while the fraction of this momentum 
carried by the struck quark is denoted $\beta$. These variables 
are related to Bjorken $x$ by $x=\beta \, x_{\spom}$. 

Diffractive interactions are often discussed 
in the framework of Regge phenomenology~\cite{regge}
in terms of  
the exchange of a `pomeron' with vacuum quantum numbers.
This interpretation in terms of a universal exchange is 
experimentally supported by the 
`proton vertex factorisation',
which holds to good approximation over much of the accessible
kinematic range at low $x_{\spom}$, whereby the dependences
on variables describing the soft interaction with the proton ($x_{\spom}$, $t$)
factorise from those related to the hard interaction 
with the virtual photon ($\beta$, $Q^2$).
Similar reactions, in which
sub-leading Reggeon and pion trajectories are exchanged, 
have a negligible cross section at the smallest $x_{\spom}$ values.

Significant progress has been made in understanding diffraction 
in terms of QCD by studying virtual photon dissociation 
in deep 
inelastic electron-proton
($ep$) scattering (diffractive DIS) at HERA. A recent 
review can be found at~\cite{review}.
As well as being sensitive to novel features of parton dynamics
in the high density, low $x$ regime, 
diffractive DIS cross sections are used to
extract diffractive parton density functions 
(DPDFs) \cite{h1lrgold,ptagging2,watt,h1lrg,h1jets}, 
an essential ingredient
in predicting many diffractive processes at the LHC and in estimating
backgrounds to more exotic processes such as central exclusive Higgs
production ($pp \rightarrow pHp$)~\cite{cep}.

Similarly to inclusive DIS, 
cross section measurements for the reaction $ep \to eXp$ 
are conventionally expressed in terms of the
reduced diffractive cross section, $\sigma_r^{D(3)}$, which is related
to the measured cross section by
\begin{eqnarray}
\frac{{\rm d} \sigma^{ep \rightarrow eXp}}{
{\rm d} \beta {\rm d} Q^2 {\rm d} x_{\spom}} = 
\frac{4\pi\alpha^2}{\beta Q^4} \ \ \left[1-y+\frac{y^2}{2}\right] \ \ 
\sigma_r^{D(3)}(\beta,Q^2,x_{\spom}) \ .
\label{sigma-2}
\end{eqnarray}
At moderate inelasticities $y$, $\sigma_r^{D(3)}$
corresponds
to the diffractive structure function $F_2^{D(3)}$ to good approximation.
In this contribution, we 
tackle the technical issue of compatibility between different
$\sigma_r^{D(3)}$ data sets through
detailed comparisons between different
measurements by the H1 and ZEUS collaborations and
take the first steps towards a combined HERA data set.
 
\section{Methods of selecting diffraction at HERA}

Experimentally, diffractive $ep$ scattering 
is characterised by the presence of a leading proton in the 
final state retaining most of the initial state proton energy, and
by a lack of hadronic activity in the 
forward (outgoing proton) direction, such that the
system $X$ is cleanly separated and $M_X$ may be measured in the central
detector components.  
These signatures have been widely exploited at HERA to select 
diffractive events by tagging the outgoing proton
in the H1 Forward Proton Spectrometer or 
the ZEUS Leading Proton Spectrometer
(proton-tagging method~\cite{ptagging4,h1fpsold,ptagging3,ptagging2,h1fps}) or
by requiring the presence of a large gap in the rapidity 
distribution of hadronic final state particles 
in the forward region (LRG method~\cite{lrgnoi,h1lrg,h1lrgold,H1:newdata}). 
In a third approach ($M_X$ method~\cite{mx,mx1,mx2,H1:newdata}), the inclusive
DIS sample is
decomposed into diffractive and non-diffractive contributions based
on their characteristic dependences on $M_X$.

The kinematic coverages of the LRG and $M_X$ methods
are limited to $x_{\spom} \lapprox 0.05$
by the need to contain the system $X$ in the central
detector components. These two
methods are equivalent for $M_X 
\rightarrow 0$, but differences are to be expected at larger
$M_X$, where the 
LRG method measures the full cross section 
from all sources at a given 
($x_{\spom}, \beta, Q^2$) 
point, whereas the
$M_X$ method involves the subtraction of a 
`non-diffractive' component. LPS and FPS
data extend to $x_{\spom} \sim 0.1$ and are therefore the
most sensitive to non-leading contributions, including
Reggeon and pion trajectory exchanges. 
Apart from the proton dissociation treatment in the H1 case
(see section~\ref{sec:h1lrg}), the cross sections measured 
by the proton-tagging and LRG methods are equivalent.  

The methods differ substantially in their dominant sources of 
systematic uncertainty. In the LRG and $M_X$ methods, 
the largest uncertainties arise from the admixture of low mass
leading baryon systems other than protons. These include 
proton excitations to low mass states as well as leading
neutrons produced via charge exchange reactions. All such 
contributions are collectively referred to here as `proton dissociation', 
$ep \rightarrow eXN$, with the baryon state
$N$ having mass $M_N$. 
Proton dissociation processes cannot always be distinguished by
the LRG and $M_X$ methods from events
in which the proton is scattered elastically.
Conversely, low-$x_{\spom}$ samples selected by the proton-tagging 
method have little or no 
proton dissociation background, 
but are subject to large uncertainties in the proton
tagging efficiency, which is strongly dependent on the proton-beam optics.
Proton spectrometers also allow a measurement 
of $t$, 
but the statistical precision is limited by their 
small acceptances.

Comparing the results from the three different methods is
a powerful test of the control over the systematics of the 
measurements. At low $x_{\spom}$, 
the ratio of results obtained by the LRG and $M_X$ methods to those from
the proton-tagging method 
can also be used to quantify the proton dissociation contributions in 
the former samples.


\section{Data sets}

A comprehensive comparison has been carried out
between recent H1 and ZEUS measurements
obtained with the three different methods.
The data sets used are as 
follows.\footnote{The comparisons here are restricted to published data
and do not yet include the precise H1 LRG and $M_X$ method results obtained
from 1999-2004 running \cite{H1:newdata}.}
\begin{itemize}
\item
Three data sets collected with the ZEUS detector  
in the years 1999 and 2000. Overlapping samples  
have been analysed with the ZEUS Leading Proton Spectrometer 
(termed {\bf ``ZEUS LPS''}, 
based on a luminosity of 32.6 pb$^{-1}$)~\cite{lrgnoi}, 
with the LRG method ({\bf ``ZEUS LRG''}, 
62.4 pb$^{-1}$)~\cite{lrgnoi} and with the 
$M_X$ method, relying on the Forward Plug Calorimeter 
({\bf ``ZEUS FPC I''}, 4.2 pb$^{-1}$~\cite{mx1} and {\bf ``ZEUS FPC II''}, 
63.4 pb$^{-1}$~\cite{mx2}).

\item
A set of data 
collected with the H1 Forward Proton Spectrometer ({\bf ``H1 FPS''},
28.4 pb$^{-1}$) \cite{h1fps} in the years 1999 and 2000.

\item
A set of data collected with the H1 detector in the years 1997, 1999 and 2000 
and analysed with the LRG method
({\bf ``H1 LRG''}, 2.0 pb$^{-1}$, 
10.6 pb$^{-1}$ and 61.6 pb$^{-1}$ for small, intermediate and large
$Q^2$, respectively)\cite{h1lrg}.

\end{itemize}

The H1 LRG and FPS samples are statistically independent and are only weakly 
correlated through systematics. 
The three ZEUS samples 
also have different dominating systematics, 
but are not statistically independent. About 75\% of events are
common to both the ZEUS LRG and ZEUS FPC II data sets and 
35\% of the ZEUS LPS events are also contained in the ZEUS LRG sample.

\section{Proton dissociation background and corrections}
\label{sec:pdiss}

\begin{wrapfigure}{r}{0.5\columnwidth}
\vspace{-0.6cm}
\centerline{\includegraphics[width=0.5\columnwidth]{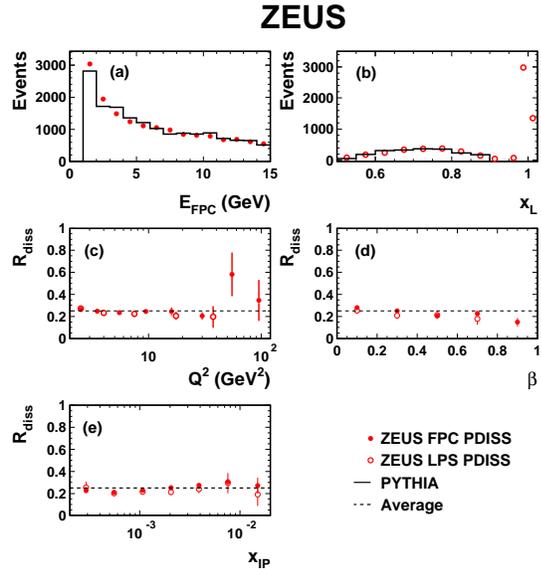}}
\vspace{-0.5cm}
\caption{(a) FPC energy and (b) LPS $x_L$ distributions
for ZEUS proton dissociation samples (see text), with
data compared to the tuned {\sc Pythia} model. (c-e) Extracted
fractions of proton dissociation events in the ZEUS LRG sample
as a function of $Q^2$, $\beta$ and $x_{\spom}$ after integration
over the other variables
\protect\cite{lrgnoi}\protect.}\label{fig:pythia}
\end{wrapfigure}

In proton dissociation processes at the lowest $M_N$, the 
dissociative system $N$ often escapes entirely 
undetected into the forward beam-pipe. As
$M_N$ increases, it becomes more likely that dissociation products are
detected in the 
instrumentation most sensitive to forward energy flow. 
The LRG and $M_X$ methods therefore do not distinguish low $M_N$
proton dissociation events from the case in which the proton is
scattered elastically. Different cross-section definitions have been
adopted, in
which the proton dissociation contribution is either subtracted
statistically, or else the quoted results are integrated over a specific
range of $M_N$.
Since understanding 
the proton dissociation contributions and the corresponding 
corrections is fundamental to comparisons between the different measurements,
a detailed discussion is presented in the following.  

In both the ZEUS LPS and the H1 FPS analyses,
the contribution from proton dissociation events
is negligible at small $x_{\spom} \lapprox 0.02$. At the largest 
$x_{\spom}$ values, it becomes kinematically possible for the 
detected leading proton to be the result of a decay of an $N^*$ 
or other proton excitation, the remaining decay products being unobserved.
This background was estimated by ZEUS
to contribute around 9\% at $x_{\spom}=0.1$, using 
the {\sc Pythia} Monte Carlo (MC) model~\cite{pythia}. 
In the H1 FPS analysis, using the {\sc Rapgap}~\cite{rapgap} implementation
of the {\sc Diffvm} 
proton dissociation model~\cite{diffvm}, it was estimated to
reach $3 \%$ at $x_{\spom} = 0.08$.


Proton dissociation contributions in the LRG and $M_X$ methods
can be controlled using dedicated proton dissociation simulations tuned
in $M_N$ regions where dissociating protons leave signals in the detectors,
and extrapolated into the $M_N$ regions where the dissociation
products are typically not detected. 
In addition to this procedure, 
both H1 and ZEUS use standard simulations of non-diffractive 
processes to control the small migrations of 
very high $M_N$ or $x_{\spom}$ events
into the measurement
region, which occur due to inefficiencies of the forward detectors.

\subsection{ZEUS LRG}
\label{pdiss:lrg}

\begin{wrapfigure}{hr}{0.5\columnwidth}
\vspace{-0.7cm}
\centerline{\includegraphics[height=.35\textheight]{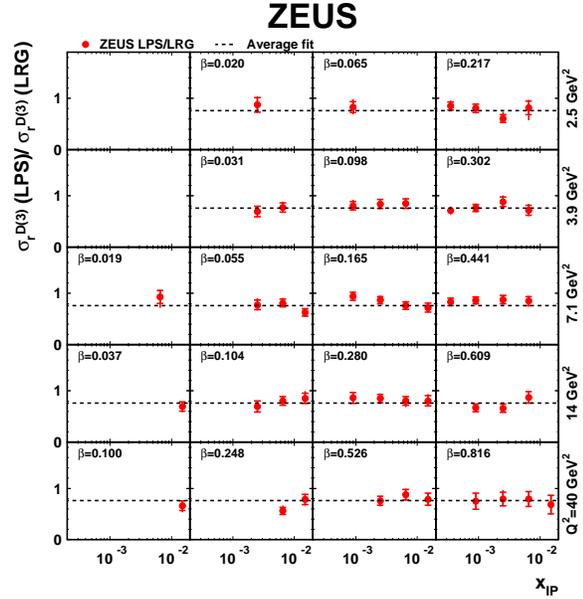}}
\vspace{-0.6cm}
\caption{The ratio of the ZEUS LPS measurement ($M_N = m_p$) to the
ZEUS LRG measurement before subtraction of proton dissociation
background 
\protect\cite{lrgnoi}\protect. 
The lines represent the average value of this ratio.
An overall normalisation uncertainty of $^{+11}_{-8} \%$ is not 
included in the errors shown.}
\label{fig:ratiozeus}
\end{wrapfigure}

In the recent ZEUS
analysis, the {\sc Pythia} simulation was tuned to proton dissociation
signals.
Two samples were selected by requiring activity either in the
forward plug calorimeter (FPC) or at relatively low proton energy in the LPS.
The samples thus include the low $M_N$ region in which proton 
dissociation products are invisible to the central detector.
The generated distributions were reweighted in $M_N$, $M_X$ and $Q^2$ 
to best describe
the energy distribution in the FPC ($E_{\rm FPC}$), and the
scattered proton energy fraction distribution ($x_L$) in the LPS.
Figures~\ref{fig:pythia}a
and~\ref{fig:pythia}b show the comparison of 
the reweighted {\sc Pythia} model with
the two proton dissociation samples as a function of these variables. 
Also shown in
figures~\ref{fig:pythia}c-e is the 
resulting estimate of the fraction of proton dissociation
events in the LRG sample as a function of $Q^2$, $\beta$ and
$x_{\spom}$. This fraction, obtained separately from the 
FPC and LPS samples, is constant at the level of $25$\%.

The ratios of cross sections extracted from the ZEUS LPS and 
LRG data (the latter uncorrected for proton dissociation background), 
are 
shown in figure~\ref{fig:ratiozeus}. 
There is no significant dependence
on
$Q^2$, $x_{\spom}$ or $\beta$, illustrating the 
low $x_{\spom}$ compatibility between the two methods.
The ratio averages to
$0.76 \pm 0.01 {\rm (stat.)} ^{+0.03} _{-0.02}
{\rm (syst.)} ^{+0.08} _{-0.05} {\rm (norm.)} $,
the last error reflecting the normalisation uncertainty of the
LPS data.
The proton dissociation
background fraction in the LRG data is thus $24 \pm 1
{\rm (stat.)} ^{+2} _{-3} {\rm (syst.)} ^{+5} _{-8}{\rm (norm.)} $\%, in
agreement with the result of the MC study, 
$25 \pm 1 {\rm (stat.)} \pm 3 {\rm (syst.)}$\% (figure~\ref{fig:pythia}). 
Unless stated otherwise, the ZEUS LRG data are corrected
by this factor in the following and thus correspond exclusively to the truly 
proton-elastic process.

\subsection{H1 LRG}
\label{sec:h1lrg}

The contribution from proton dissociation 
in the H1 LRG analysis is constrained through the 
{\sc Diffvm} MC~\cite{diffvm} model, normalised 
using the response to large $M_N$ events leaving 
signals in the forward and central detector 
components \cite{h1lrg,carrie}.

\begin{figure}[ht]
  \vspace*{0.3cm}
  \includegraphics[height=.20\textheight]{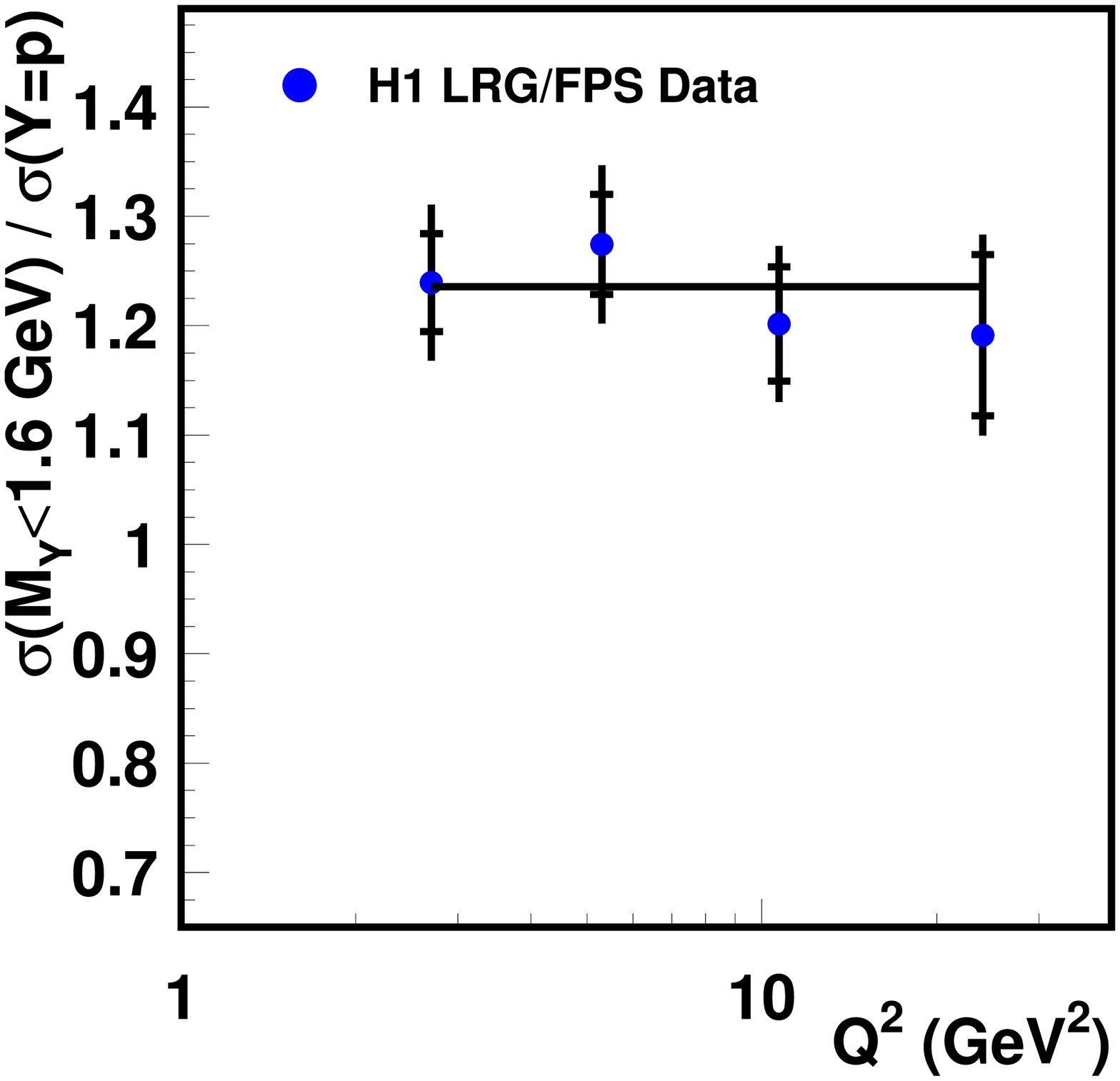}
  \includegraphics[height=.20\textheight]{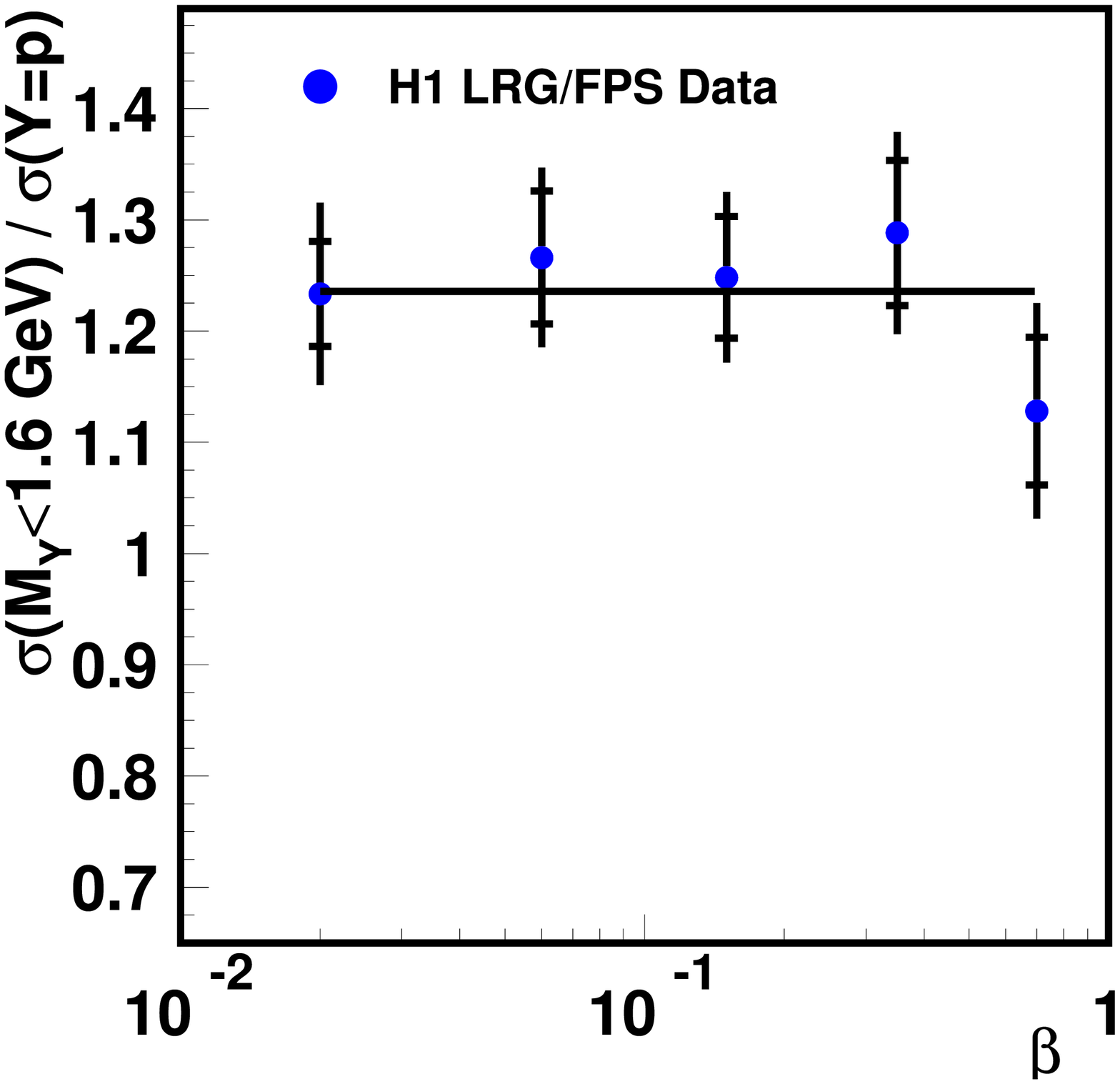}
  \includegraphics[height=.20\textheight]{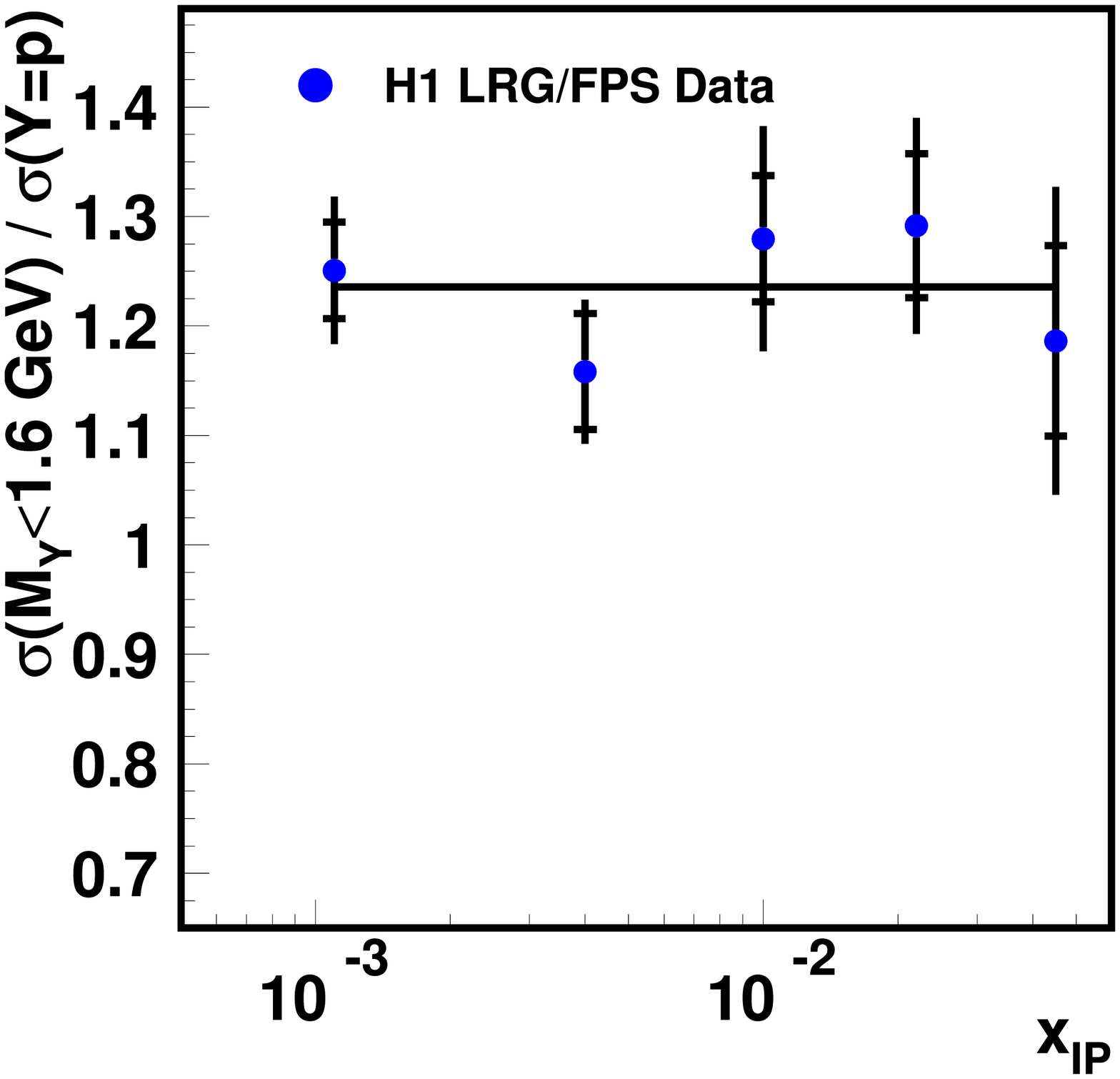}
\vspace*{-0.5cm}
  \caption{The ratio of the H1 LRG measurement 
(corrected to $M_N < 1.6 \ {\rm GeV}$) to the 
H1 FPS measurement ($M_N = m_p$), after integration over the 
variables not shown in each case 
\protect\cite{h1fps}\protect. The lines represent a fit
to the data assuming no dependence on any of the variables.
An overall normalisation uncertainty of $13 \%$ is not 
included in the errors shown.}
  \label{fig:ratioh1}
\end{figure}

The data are corrected using {\sc Diffvm} to $M_N < 1.6$~GeV.
The H1 LRG data are then compared 
with the H1 FPS measurement, in order to extract 
the proton dissociation cross section with $M_N < 1.6 \ {\rm GeV}$
directly from the data. 
The ratio of the two measurements, 
after projection onto the $Q^2$, $x_{\spom}$ and $\beta$ axes,
is shown in figure~\ref{fig:ratioh1}. There is no evidence for
any dependence on any of the kinematic variables.
as expected in the framework of proton vertex factorisation. The
average value of the ratio is 
$1.23 \pm \ 0.03 \ {\rm (stat.)} \ \pm \ 0.16 \ {\rm (syst.)}$, 
the largest uncertainty arising from the FPS efficiency.
The result is in
good agreement with the 
{\sc Diffvm} estimate of $1.15^{+0.15}_{-0.08}$. 
The data and {\sc Diffvm} ratios translate into 
proton dissociation 
background fractions of $19 \ \%$ 
and $13 \ \%$, respectively, consistent within the uncertainties.
The similarity between the proton dissociation fractions in the raw 
H1 and ZEUS LRG selections is to be expected given the
similar forward detector acceptances of the two 
experiments. 

\subsection{ZEUS FPC}
\label{pdiss:mx}

The proton dissociation treatment is also critical in the
$M_X$ method, where
the diffractive contribution is separated from
the non-diffractive
component in a fit to the inclusive $\ln M_X^2$ distribution.
Proton dissociation events with sufficiently large $M_N$ for 
dissociation products to reach the FPC and central detectors
lead to a reconstructed $M_X$
value which is larger than the actual photon dissociation mass. The
resulting distortion of the $\ln M_X^2$ distribution affects the 
diffractive contribution extracted in the fit if corrections are not made.
According to the {\sc Sang} MC model,
the $N$ system contaminates
the $M_X$ reconstruction for $M_N > 2.3$ GeV 
on average~\cite{heuijin_thesis}, 
and events in this $M_N$ range are therefore subtracted using {\sc Sang} 
before the $\ln M_X^2$ distribution is decomposed. 
The upper $M_N$ cut in the {\sc Sang} sample is defined by 
$(M_N/W)^2 < 0.1$, which leads to a variation of the 
subtracted fraction of events with $W$, 
the centre-of-mass energy of the photon-proton system.
This contrasts with the LRG method, where
MC studies confirm that the 
rapidity gap requirement efficiently eliminates
proton dissociation at large $M_N$, the remaining 
fractional low $M_N$ contribution being independent of kinematics
to good approximation (figures~\ref{fig:ratiozeus} and \ref{fig:ratioh1}).

Despite these difficulties, there is
acceptable agreement between the ZEUS FPC 
data and the ZEUS LRG measurement. A global fit 
comparing the normalisations of the two data sets (after correcting
the LRG data to $M_N = m_p$) yields a
normalisation factor of 
0.83 $\pm$ 0.04 to be applied to the ZEUS FPC results. 
This factor is comaptible with 
with expectations for the residual proton dissociation 
contribution based on the MC studies 
in sections~\ref{pdiss:lrg} and~\ref{sec:h1lrg}.

\begin{figure}[ht]
\vspace*{-0.3cm}
  \includegraphics[height=.36\textheight]{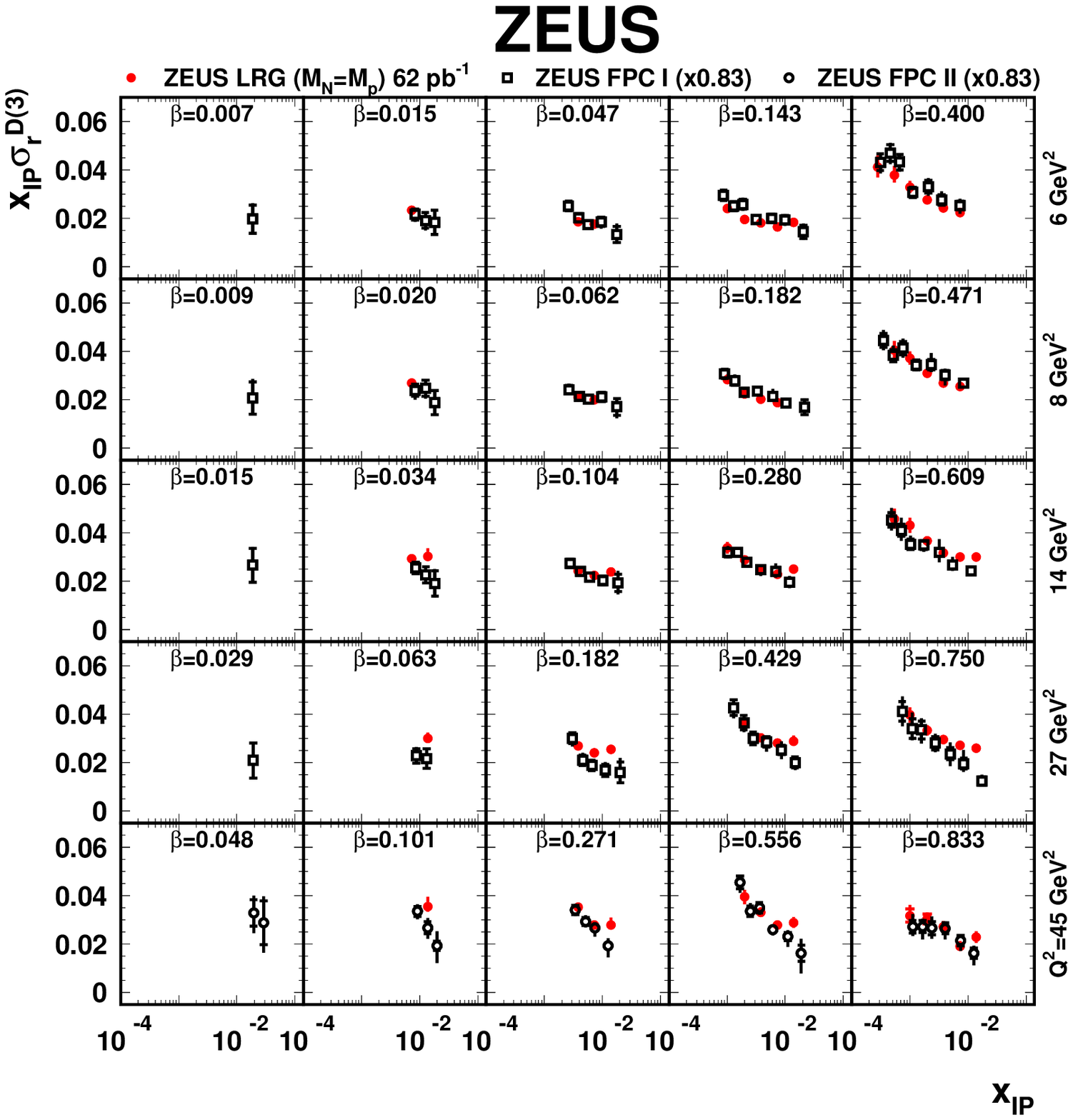}
  \hspace{-0.6cm}
  \includegraphics[height=.36\textheight]{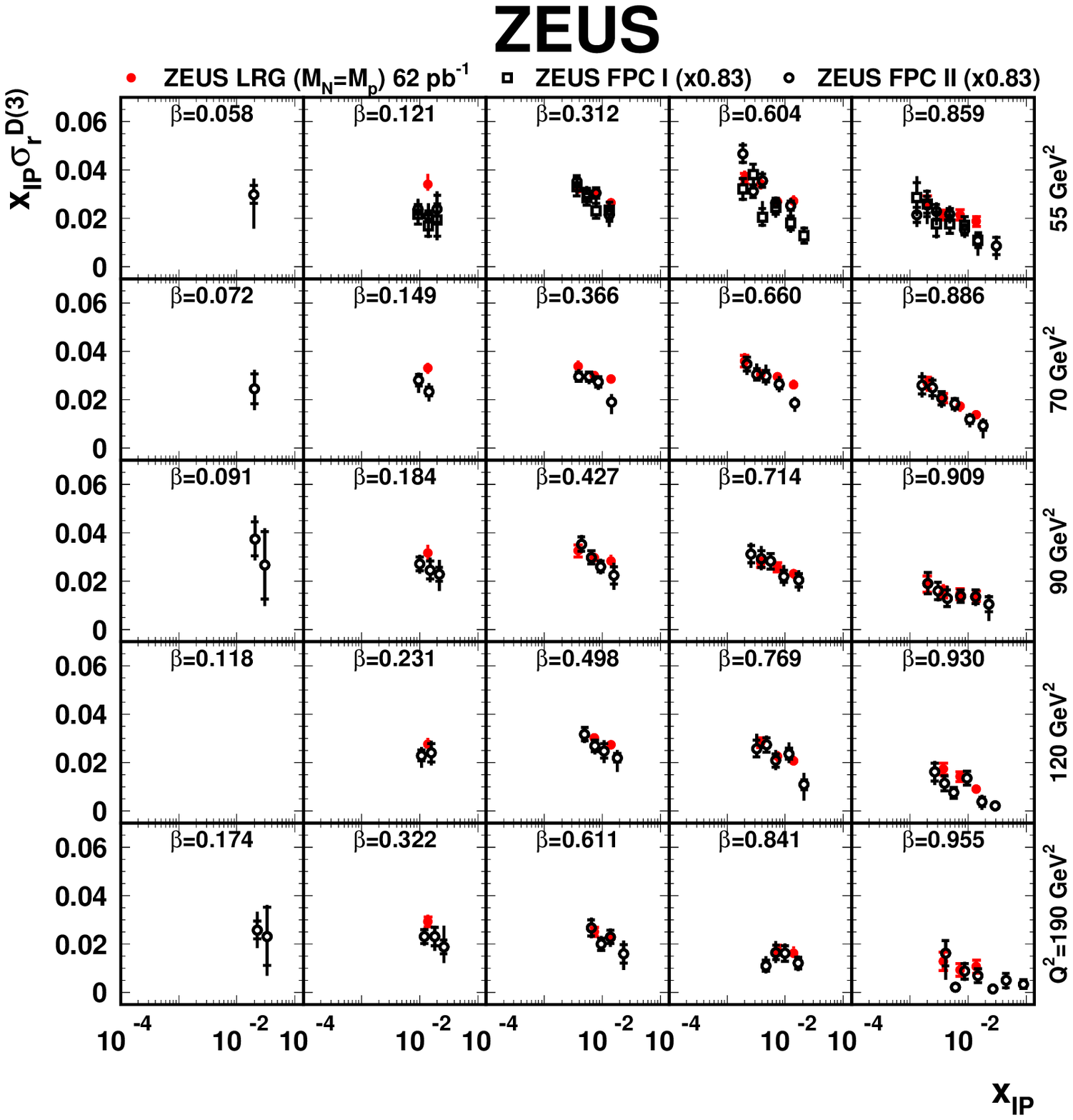}
\vspace*{-0.3cm}
  \caption{Comparison between the ZEUS $M_X$ method (`FPC I' and `FPC II')
and ZEUS LRG method data 
\protect\cite{lrgnoi}\protect. As explained in the text,
the $M_X$ method data are scaled by a constant 
factor of 0.83 to account for proton dissociation contributions with 
$M_N < 2.3$ GeV.}
  \label{fig:comp_mx}
\end{figure}

\section{Cross section comparisons}
\label{sec-comp}

Due to their differing $M_N$ coverages, the $\sigma_r^{D(3)}$ 
measurements from the different
data sets are not directly comparable. However, 
assuming the factorisation of the $M_N$ dependence
which is suggested in 
the data, varying the $M_N$ range should introduce only global 
normalisation differences, which can be estimated using the 
proton dissociation simulations. 


\pagebreak

\subsection{Comparison between LRG and {\boldmath $M_X$} methods}
\label{lrgvmx}

\begin{wrapfigure}{r}{0.5\columnwidth}
\vspace*{-0.6cm}
\centerline{\includegraphics[height=.34\textheight]{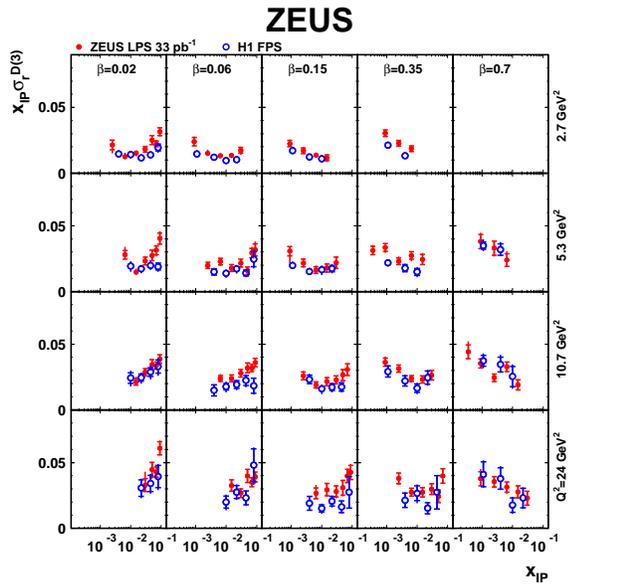}}
\vspace*{-0.5cm}
\caption{Comparison between ZEUS LPS and H1 FPS 
measurements \protect\cite{lrgnoi}\protect.
Normalisation uncertainties of $\pm 10 \%$ (H1) and $^{+11 \%}_{-7 \%}$
(ZEUS) are not shown.}
\label{fig:comp_lpsfps}
\end{wrapfigure}

ZEUS cross section measurements 
obtained with the LRG and $M_X$ methods 
are compared in figure~\ref{fig:comp_mx}.
The LRG data  
are corrected to $M_N=M_p$ as described in section~\ref{pdiss:lrg}
and the 
relative normalisation factor of 0.83 (section~\ref{pdiss:mx})
is applied to the ZEUS FPC data to account for
residual proton dissociation.
The overall agreement 
between the two measurements 
is good, apart from some differences at large $x_{\spom}\gtrsim0.01$. 
The $Q^2$ dependence of the $M_X$ method data is also 
slightly weaker than that of the LRG data.



\subsection{Comparison between ZEUS LPS and H1 FPS measurements}
\label{lpscomp}

The ZEUS LPS and H1 FPS data are compared  
in figure~\ref{fig:comp_lpsfps}. For this comparison,
the ZEUS results are extracted at the same $\beta$ and $Q^2$ values as 
H1 and are therefore not affected by extrapolation uncertainties.  
The shape agreement is satisfactory and the overall normalisation 
discrepancy of around $10$\% lies within the large
combined normalisation uncertainty of around $14\%$.



\subsection{Comparison between ZEUS and H1 LRG measurements}
\label{lrgcomp}

The ZEUS LRG data are extracted at the H1 $\beta$ and $x_{\spom}$ 
values, but at different $Q^2$ values.
In order to match the $M_N~<~1.6$\,GeV range of the H1 data, 
a global factor of $0.91 \pm 0.07$, estimated with {\sc Pythia},  
is applied to the ZEUS LRG data in place of the correction
to an elastic proton cross section.
After this procedure, the ZEUS data remain higher than those of H1 
by $13\%$ on average, as estimated with a global fit 
comparing the normalisations of the two data
sets for 
$Q^2 > 6$\,GeV$^2$. This normalisation discrepancy is
similar to that between the H1 FPS and the ZEUS LPS data sets.
It is in line with the errors due to the
$8\%$ uncertainty on the proton dissociation 
correction in the ZEUS LRG data and the $7\%$ combined relative 
normalisation uncertainty between the two LRG data sets.

\begin{figure}[p] \unitlength 1mm
  \begin{center}
    \begin{picture}(100,200)
      \put(55,-5){\epsfig{file=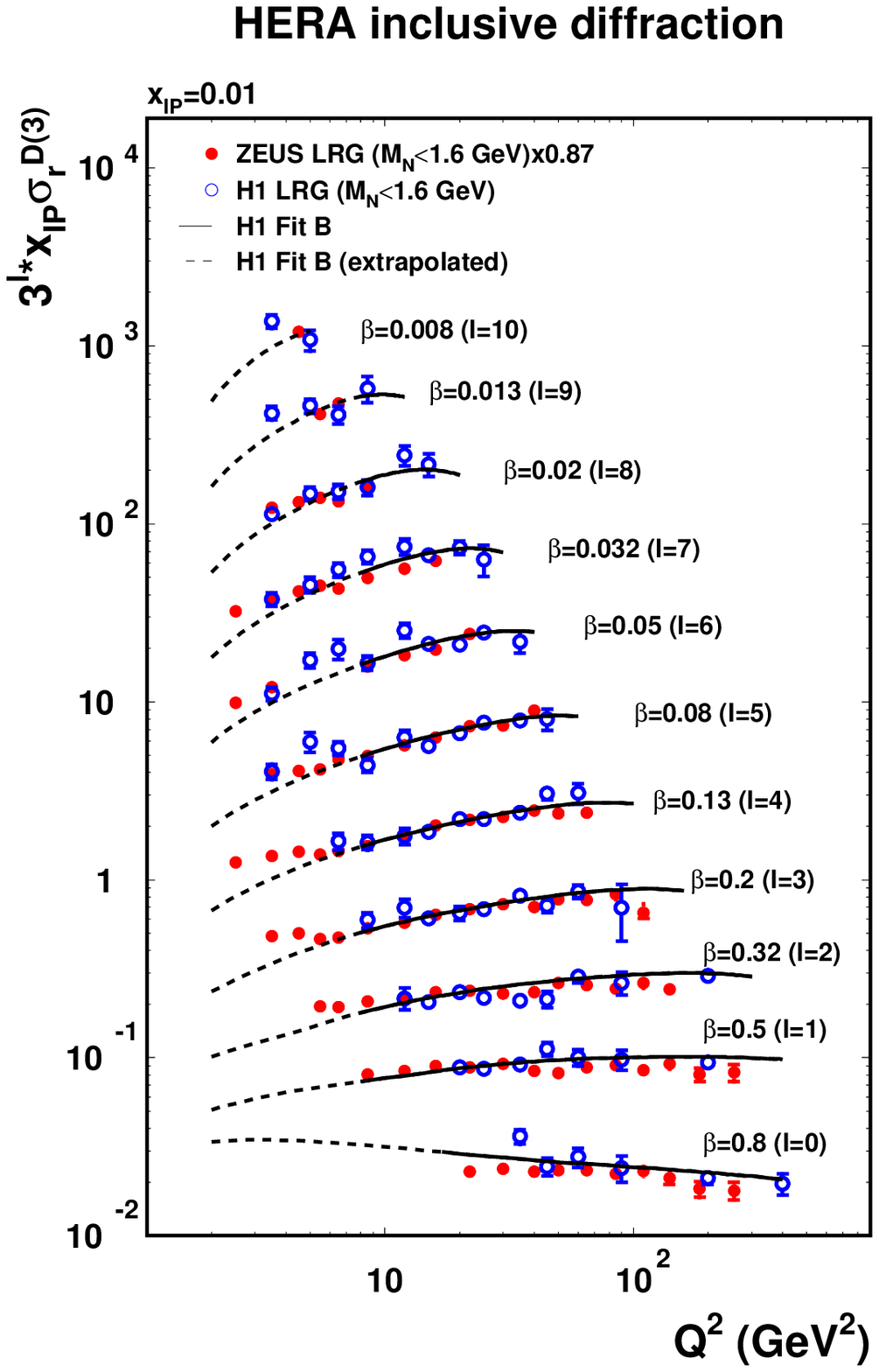,width=0.5\textwidth}}
      \put(55,105){\epsfig{file=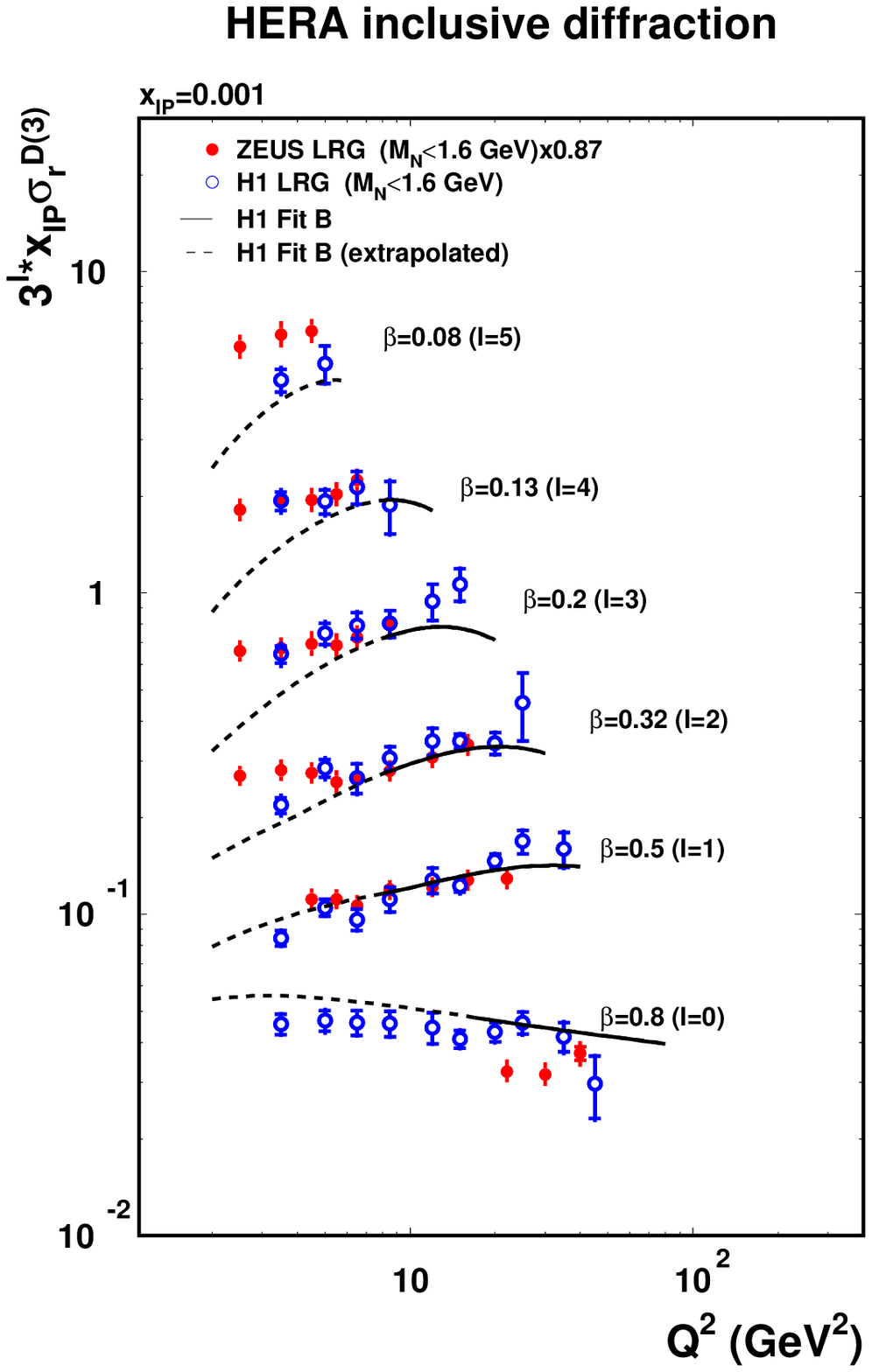,width=0.5\textwidth}}
      \put(-25,-5){\epsfig{file=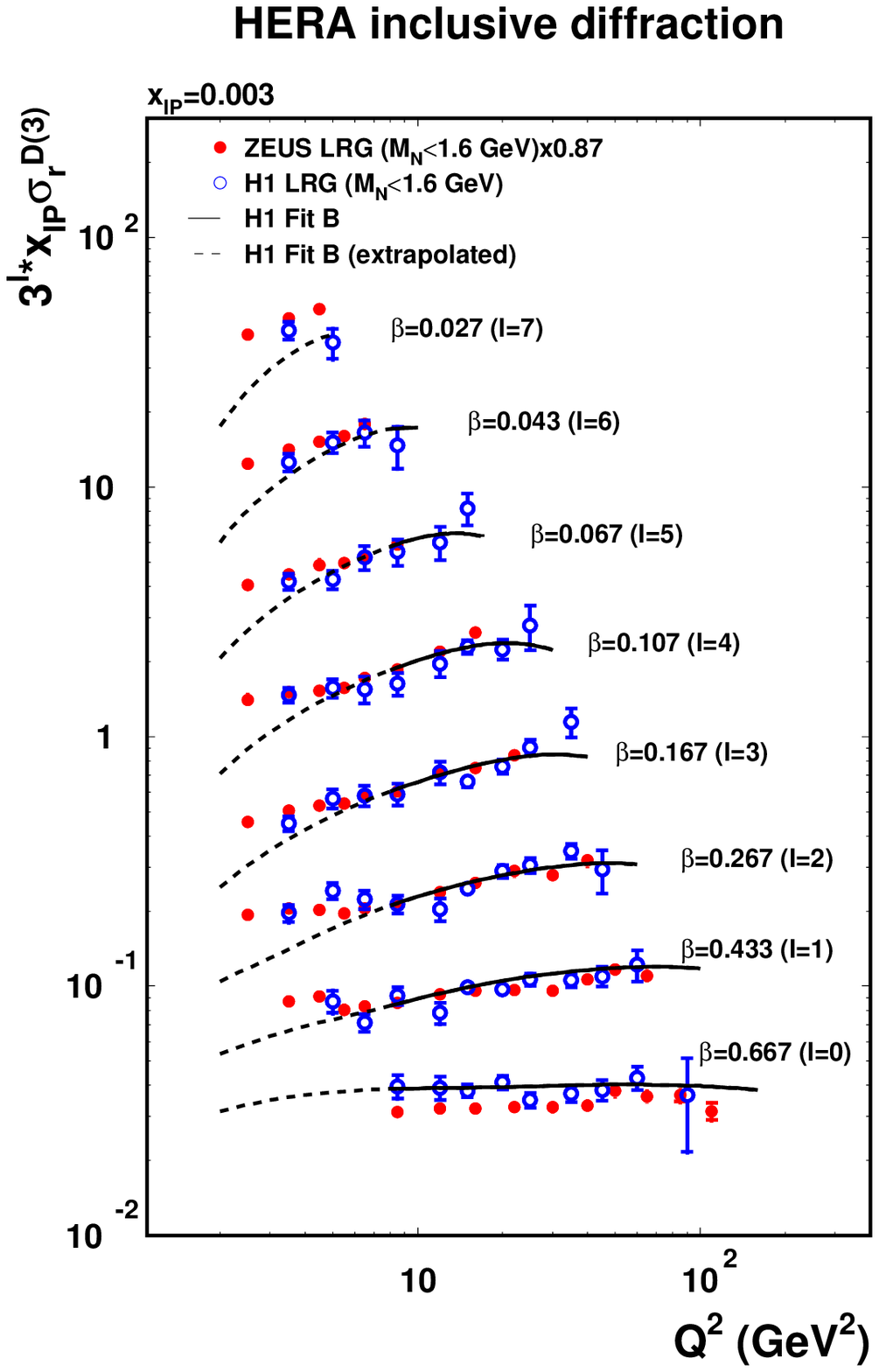,width=0.5\textwidth}}
      \put(-25,105){\epsfig{file=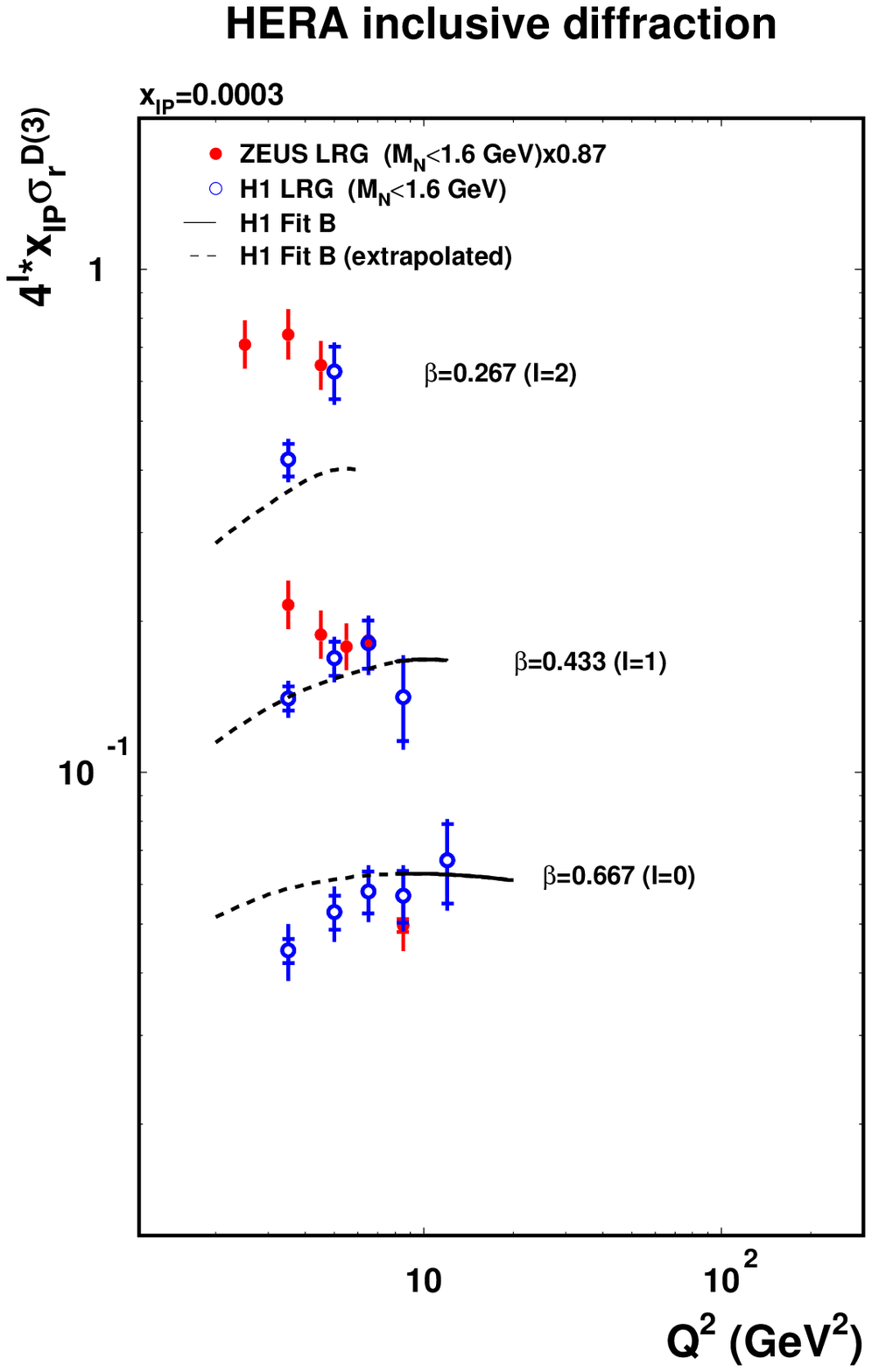,width=0.5\textwidth}}
    \end{picture}
  \end{center}
  \caption[]{Comparison between the H1 and ZEUS LRG measurements 
after correcting both data sets to $M_N  < 1.6 \ {\rm GeV}$ 
and applying a further
scale factor of 0.87 (corresponding to the average normalisation difference)
to the ZEUS data. The measurements are compared with the results
of the 'H1 Fit B' DPDF extraction, which was based on
the H1 data shown. Further H1 data at $x_{\spom} = 0.03$
are not shown.}
\label{fig:comp_lrg}
\end{figure}

In figure~\ref{fig:comp_lrg}, the ZEUS results are
scaled by a factor $0.91 \times 0.87$ (the factor $0.87~=~1-0.13$  
normalising the ZEUS to the H1 data) and compared with the 
H1 LRG measurement. An excellent agreement between the $Q^2$ 
dependences is revealed throughout most of the phase space. There are small 
deviations between the $\beta$ dependences of the
two measurements at the highest and lowest $\beta$ values. 
The results of the `H1 Fit B' NLO QCD DPDF fit to the 
H1 LRG data~\cite{h1lrg} is 
also shown. It gives a good description of the data at large $Q^2$.
However, the extrapolation beyond the fitted region 
($Q^2 \geq 8.5 \ {\rm GeV^2}$) undershoots
the precise new ZEUS low $Q^2$ LRG data, confirming the observation
in \cite{h1lrg} that a standard DGLAP fit to the lowest $Q^2$ data
is problematic.
 

\section{A First Combination of Data Sets}

For easy future consumption at the LHC and elsewhere, it is desirable
to combine  
the various H1 and ZEUS diffractive DIS measurements
into a single easily digestible HERA data set. Here we take the
first steps towards this goal, by making a simple error-weighted average of 
the H1 and ZEUS LRG data sets, ignoring correlations between the data
points due to the systematic errors. LPS and $M_X$ method data are not
considered at this stage.
For the purpose of this exercise, the ZEUS normalisation is fixed to that
of H1 as described in section~\ref{lrgcomp} and shown in 
figure~\ref{fig:comp_lrg}. The normalisation of the combined data
thus has an uncertainty beyond the $10 \%$ level.
Combinations can only meaningfully be made
where there is basic agreement between the 
different measurements. Since this is not always the case at the
lowest $x_{\spom}$ values, we restrict
the averaging to the $x_{\spom} = 0.003$ and $x_{\spom} = 0.01$ data.
The combinations are performed throughout the measured $Q^2$ range, 
including the $Q^2 < 8.5 \ {\rm GeV}^2$ region, 
beyond the range of the `H1 Fit B' parameterisation which  
is compared with the data.

To account for the differences between the 
$Q^2$ binning choices, H1 data points are adjusted to the ZEUS 
$Q^2$ values by applying small correction factors calculated using 
the `H1 Fit B' parameterisation. Where both collaborations then 
have measurements at a given ($Q^2, x_{\spom}, \beta$) point, a 
simple weighted average is taken, 
using the quadratic sum of statistical and systematic uncertainties
for each experiment, excluding normalisation uncertainties.
At ($Q^2, x_{\spom}, \beta$) points where only one experiment 
has a measurement, that result is carried directly into the combined
data set.

\begin{figure}[ht]
\vspace*{-0.3cm}
  \includegraphics[height=.5\textheight]{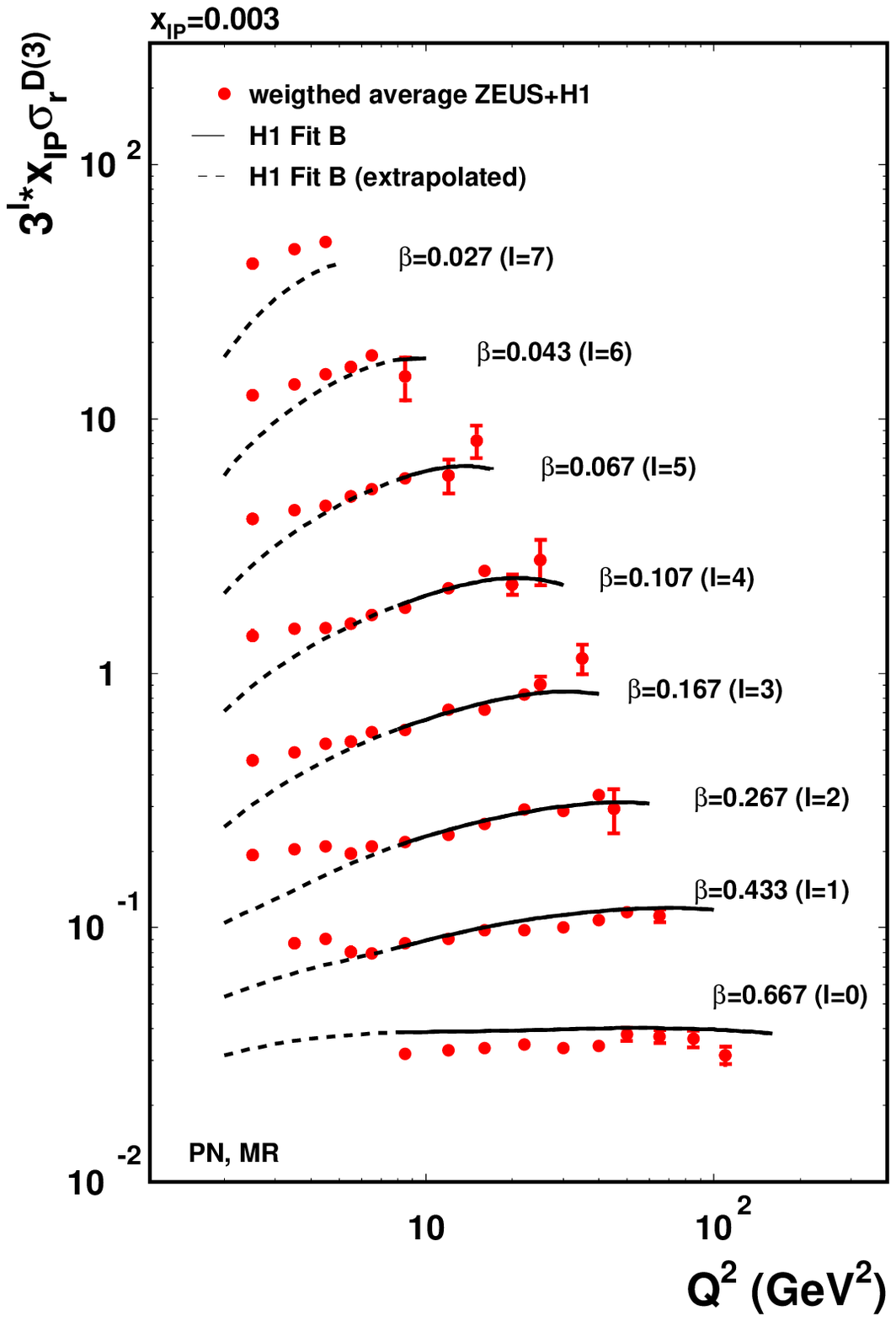}
  \hspace{-0.6cm}
  \includegraphics[height=.5\textheight]{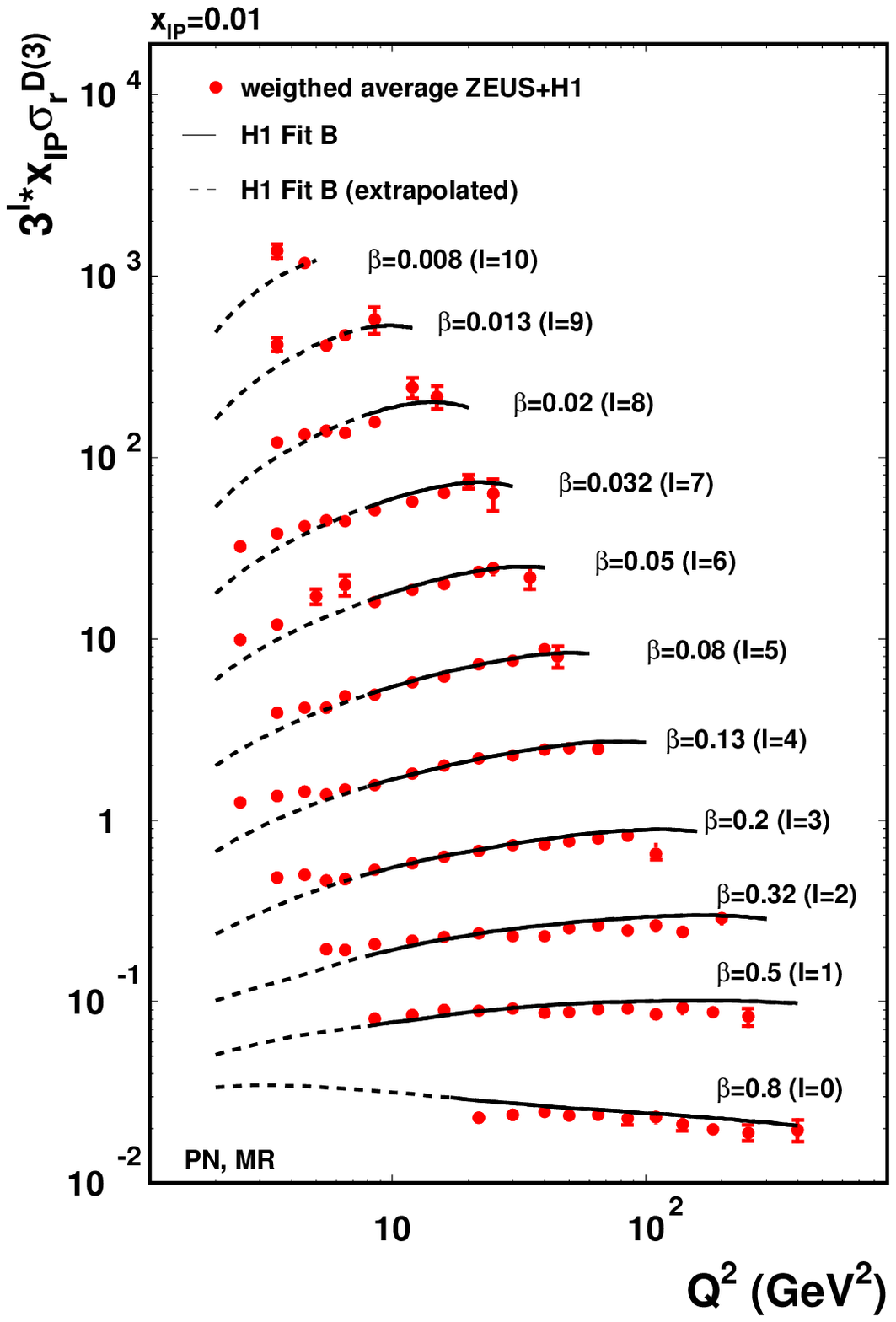}
\vspace*{-0.3cm}
  \caption{Combination of the H1 and ZEUS LRG data following the
procedure described in the text. The global normalisation is fixed to that
of the H1 measurement, in order most easily to compare the data with the
'H1 Fit B' DPDF results.}
  \label{fig:average}
\end{figure}

The results of this averaging procedure\footnote{The 
authors take full responsibility for this 
combined data set; it is not an official H1/ZEUS result.} 
are shown in figure~\ref{fig:average}. 
They are indicative of the
sort of precision which is achievable through combinations, with many data
points having errors at the $3-4 \%$ level, excluding the normalisation
uncertainty. At $x_{\spom} = 0.01$ the combined data agree well with the
`H1 Fit B' DPDF parameterisation in its region of validity. 
At $x_{\spom} = 0.003$ the $Q^2$ dependences
are also in good agreement with the parameterisation in the 
$\beta$ and $Q^2$ region of the fit, 
with the exception of the highest $\beta$ value, where the average
is pulled towards the more precise ZEUS data. 

More sophisticated
averaging methods may be used in the future, for example that \cite{glazov} 
developed to perform similar combinations of
inclusive HERA data, with a full systematic error treatment. 
No attempt has yet been made to extract DPDFs from the combined data. 
Based on the combined $\sigma_r^{D(3)}$ and its $Q^2$ dependence
shown here, 
no significant conflict is expected with the quark or gluon densities 
of existing DPDF sets such as `H1 Fit B' 
in the bulk of the phase space. However, 
small modifications are likely to be necessary to the quark densities 
at small and large $\beta$ values and understanding the region
$Q^2 \lapprox 8.5 \ {\rm GeV^2}$ remains a challenge.


\section{Summary}

H1 and ZEUS diffractive DIS data obtained by various methods with 
very different systematics have been compared in detail. 
All
measurements are broadly consistent in the shapes of the distributions.
The comparisons
between proton tagging and LRG method data internally to the two collaborations
give compatible results on the proton dissociation contributions in the 
raw LRG selections. There is  a global normalisation difference at the 
13\% level 
between the LRG measurements of the
two experiments, which is a little beyond 
one standard deviation in the combined normalisation uncertainty. A similar
difference is visible between the normalisations of the 
H1 and ZEUS proton tagged data.

A first step has been taken towards combining the two sets of LRG data,
by arbitrarily fixing the normalisation to that of the H1 data set and 
ignoring correlations within the systematic uncertainties in obtaining
weighted averages. The results hint at the precision which might be obtained
in the future with a more complete procedure. 

\pagebreak

\bibliographystyle{heralhc} 
{\raggedright

\bibliography{heralhc}

\providecommand{\etal}{et al.\xspace}
\providecommand{\coll}{Coll.}
\catcode`\@=11
\def\@bibitem#1{%
\ifmc@bstsupport
  \mc@iftail{#1}%
    {;\newline\ignorespaces}%
    {\ifmc@first\else.\fi\orig@bibitem{#1}}
  \mc@firstfalse
\else
  \mc@iftail{#1}%
    {\ignorespaces}%
    {\orig@bibitem{#1}}%
\fi}%
\catcode`\@=12
\begin{mcbibliography}{10}

\bibitem{heralhc}
Proceedings of the 2nd, 3rd and 4th HERA-LHC Workshops,
\newblock in litt., eds. de Roeck, Jung (2009);
  \verb+http://www.desy.de/~heralhc/+{}\relax
\relax
\bibitem{zeus:obs}
ZEUS Collaboration, M. Derrick et al.,
\newblock Phys. Lett.{} {\bf B 315},~481~(1993)\relax
\relax
\bibitem{h1:obs}
H1 Collaboration, T. Ahmed et al.,
\newblock Nucl. Phys.{} {\bf B 429},~477~(1994)\relax
\relax
\bibitem{regge}
P.D.B.~Collins,
\newblock {\em An Introduction to {Regge} Theory and High Energy Physics}.
\newblock Cambridge University Press, Cambridge, 1977\relax
\relax
\bibitem{review}
M.~Arneodo and M.~Diehl,
\newblock Procedings of the 1st HERA-LHC Workshop, eds. de Roeck, Jung,
  CERN-2005-014, hep-ph/0511047{}\relax
\relax
\bibitem{h1lrgold}
H1 Collaboration, C. Adloff et al.,
\newblock Z. Phys.{} {\bf C 76},~613~(1997)\relax
\relax
\bibitem{ptagging2}
ZEUS Collaboration, S. Chekanov et al.,
\newblock Eur. Phys. J.{} {\bf C 38},~43~(2004)\relax
\relax
\bibitem{watt}
A. Martin, M. Ryskin, G. Watt,
\newblock Eur. Phys. J.{} {\bf C 44},~69~(2005)\relax
\relax
\bibitem{h1lrg}
H1 Collaboration, A. Aktas et al.,
\newblock Eur. Phys. J.{} {\bf C 48},~715~(2006)\relax
\relax
\bibitem{h1jets}
H1 Collaboration, A. Aktas et al.,
\newblock JHEP 0710:042{}~(2007)\relax
\relax
\bibitem{cep}
FP420 R\&D Collaboration,
\newblock hep-ex/0806.0302{}~(2008)\relax
\relax
\bibitem{ptagging4}
ZEUS Collaboration, J. Breitweg et al.,
\newblock Eur. Phys. J.{} {\bf C 1},~81~(1997)\relax
\relax
\bibitem{h1fpsold}
H1 Collaboration, C. Adloff et al.,
\newblock Eur. Phys. J.{} {\bf C 6},~587~(1999)\relax
\relax
\bibitem{ptagging3}
ZEUS Collaboration, S. Chekanov et al.,
\newblock Eur. Phys. J.{} {\bf C 25},~169~(2002)\relax
\relax
\bibitem{h1fps}
H1 Collaboration, A. Aktas et al.,
\newblock Eur. Phys. J.{} {\bf C 48},~749~(2006)\relax
\relax
\bibitem{lrgnoi}
ZEUS Collaboration, S.~Chekanov et al.,
\newblock DESY 08-175, submitted to Nucl. Phys. {\bf B}{}\relax
\relax
\bibitem{H1:newdata}
H1 Collaboration,
\newblock H1prelim-06-014{}~(presented at DIS2006, E. Sauvan)\relax
\relax
\bibitem{mx}
ZEUS Collaboration, S.~Chekanov et al.,
\newblock Eur.~Phys. J.{} {\bf C~25},~69~(2002)\relax
\relax
\bibitem{mx1}
ZEUS Collaboration, S.~Chekanov et al.,
\newblock Nucl.~Phys.{} {\bf B~713},~3~(2005)\relax
\relax
\bibitem{mx2}
ZEUS Collaboration, S.~Chekanov et al.,
\newblock Nucl.~Phys.{} {\bf B~800},~1~(2008)\relax
\relax
\bibitem{pythia}
T. Sj\"ostrand, L. L\"onnblad and S. Mrenna,
\newblock hep-ph/0108264{}~(2001)\relax
\relax
\bibitem{rapgap}
H.~Jung,
\newblock Comput. Phys. Commun.{} {\bf 86},~147~(1995)\relax
\relax
\bibitem{diffvm}
B.~List and A.~Mastroberardino,
\newblock Proceedings of the Workshop on MC Generators for HERA Physics{},~p.
  396~(1999)\relax
\relax
\bibitem{carrie}
C. Johnson,
\newblock {\em Ph.D. Thesis, University of Birmingham}.
\newblock 2002\relax
\relax
\bibitem{heuijin_thesis}
H. Lim,
\newblock {\em Ph.D. Thesis, The Graduate School, Kyungpook National
  University, Taegu (Republic of Korea)}.
\newblock Unpublished, 2002\relax
\relax
\bibitem{glazov}
A. Glazov,
\newblock Proceedings of DIS 2005, Madison, USA, eds Smith, Dasu,{}~(AIP
  2005)\relax
\relax
\end{mcbibliography}
}
\end{document}